\documentclass[pre,aps,showpacs,byrevtex,amsmath,amssymb,preprint]{revtex4}

\usepackage{times}
\usepackage{epsf}
\usepackage{graphicx}

\usepackage{dcolumn}
\usepackage{bm}

\begin{document}
\draft

\title{Nonlinear response of the magnetophoresis in inverse ferrofluids}

\author{Y. C. Jian$^{1,2}$, Y. Gao$^1$ and J. P. Huang$^1$\footnote{Electronic address: jphuang@fudan.edu.cn}}
\address{$^1$Surface Physics Laboratory (National Key Laboratory)
and Department of Physics, Fudan University, Shanghai 200433
\\
$^2$Department of Physics, National Tsing-Hua University,
Hsinchu300}

\begin{abstract}
Taking into account the structural transition and long-range
interaction (lattice effect), we resort to the Ewald-Kornfeld
formulation and developed Maxwell-Garnett theory for uniaxially
anisotropic suspensions to calculate the effective permeability of
inverse ferrofluids. And we also consider the effect of volume
fraction to the magnetophoretic force on the nonmagnetic spherical
particles submerged in ferrofluids in the presence of nonuniform
magnetic field. We find that the coupling of ac and dc field case
can lead to fundamental and third harmonic response in the effective
magnetophoresis and changing the aspect ratio in both prolate and
oblate particles can alter the harmonic and nonharmonic response and
cause the magnetophoretic force vanish.
\end{abstract}

 \maketitle

\section{Introduction}

Magnetic particles are used to label and manipulate biomaterials
such as cells, enzymes, proteins, DNA, to transport therapeutic
drugs in the applications such as bioseparation, immunoassays, drug
delivery, and to separate red and white blood cells~\cite{furlani}.
Magnetizable particles experience a force in a nonuniform magnetic
field caused by the magnetic polarization. This phenomenon, called
magnetophoresis $(MAP)$ has been exploited in a variety of
industrial and commercial process for separation and beneficiation
of solids suspended in liquids ~\cite{1}. Virtually, almost all
practical magnetic separation technologies exploit the
magnetophoretic forces. Such as, belt and drum devices,
high-gradient magnetic separation ($HGMS$) systems~\cite{Watson}
etc..

In recent years, inverse ferrofluids with nonmagnetic colloidal
particles (or magnetic holes) suspended in ferrofluids have drawn
much attention for their potential industrial and biomedical
applications ~\cite{2,3,4,5}. The size of the nonmagnetic particle
(such as polystyrene particle) is about 1$\sim$100$\mu$m.
Ferrofluids, also know as magnetic fluid, is a colloidal suspension
of single-domain magnetic particles, with typical dimension around
10nm, dispersed in a liquid carrier ~\cite{2}. Because the external
magnetic field will induce magnetic dipole moment in the nonmagnetic
particles, when the field intensity increases, the inverse
ferrofluids will solidify to crystals due to dipole interactions,
which is similar to ER and MR behavior. Since the nonmagnetic
particles are much larger than the magnetic fluid nanoparticles, the
magnetic fluid is treated as a one-component continuum with respect
to the larger nonmagnetic particles ~\cite{7,8}. Recently, Jian,
Huang and Tao first theoretically suggested that the ground state of
inverse ferrofluids is body-centered tetragonal $(bct)$ lattice
~\cite{9} , just as the ground states of electrorheological fluids
and magnetorheological fluids ~\cite{10,11}. Motivated by this
study, we propose an alternative structural transition from the
$bct$ to the face-centered cubic ($fcc$), through the application of
external magnetic field. In this work, we will investigate the
magnetophoretic force on the particles during the phase transition
in the crystal system considering the effects mentioned above.

Finite-frequency responses of nonlinear dielectric composite
materials have attracted great attention in both research and
industrial applications during the last two decades~\cite{Bergman}.
In particular, when a composite containing linear permeability
particles embedded in a nonlinear permeability host medium is
subjected to a sinusoidal alternating magnetic, the magnetic
response will generally consist of frequencies of higher order
harmonics under ac field~\cite{Levy,Hui,Huii,Gu,Huang,Wei}. The main
aim of the present paper is to study the the effects of the
geometric shape and volume fraction of the particles on the
nonlinear ac responses (harmonics) of the crystal system, then the
magnetophoresis of the particles, considering the local lattice
effect. We shall use the Edward-Kornfeld formulation~\cite{12,13,14}
to derive the local magnetic fields and induced dipole moments in
inverse ferrofluids and then perform the perturbation expansion
method to obtain the fundamental and higher harmonics in two cases:
single ac and dc-ac magnetic field.

Then, we apply a nonuniform magnetic field to investigate the
magnetophoresis of the nonmagnetic particles submerged in the
ferrofluids, by taking into account the effect of structural
transition and long-range interaction (lattice effect). And we also
consider the effect of volume fraction to the magnetophoretic force
exerted on the nonmagnetic particles.

This paper is organized as follows. In Sec. II, we introduce the
nonlinear characteristics in the ferrofluids and the theory of
magnetophoresis. By using the Ewald-Kornfeld formulation and the
well-known developed Maxwell-Garnett theory for uniaxially
anisotropic suspensions, we calculate the effective permeability of
the inverse ferrofluids for different harmonic response under
time-varying magnetic field in Sec. III. Then we investigate the
nonlinear response of magnetophoretic force on the nonmagnetic
particles in ferrofluids. This paper ends with a discussion and
conclusion in Sec. IV.

\newpage
\section{Theory and Formalism}\label{}

\subsection{Nonlinear characteristics in the ferrofluids}\label{}

In the standard literature, we often consider the magnetic induction
$\textbf{B}$ to be proportional with the field strength
$\textbf{H}_0$, and their relation is written as
$\textbf{B}=\mu\textbf{H}_0$, where $\mu$ is the linear
permeability. However, the simple relation is only valid at weak
intensities. When in the real situation, especially under a stronger
field, nonlinearities are introduced as $\textbf{B}=\mu
\textbf{H}_0+\xi|H_0|^2\textbf{H}_0+\eta|H_0|^4\textbf{H}_0+...$,
where $\xi$,$\eta$ are the third-order and fifth-order nonlinear
coefficient and $\eta|H_0|^4\ll\xi|H_0|^2\ll\mu$. Here we assume the
nonlinearity is not strong and consider only the lowest-order
nonlinearity for simplicity. Thus after dropping the symbol of
absolute value in the equation above, the nonlinear permeability
$\tilde{\mu}$ is obtained:

\begin{equation}
\tilde{\mu}=\mu_1+\xi_1 H_0^2
\end{equation}

For inverse ferrofluids, the peameability $\tilde{\mu}$ of host
medium ferrofluid can be derived in the form as
$\tilde{\mu}=\mu_1+\xi_1 H_0^2$ when the magnetic particles in the
ferrofluid is nonlinear to the external field, as analyzed below. As
shown in Fig.(1), considering the magnetic particles with nonlinear
$\tilde{\mu}_p$ in the ferrofluid , the local magnetic field
$\textbf{H}_1$ inside the particle induced by external field
$\textbf{H}_0$ is give by

\begin{equation}
\textbf{H}_1=\frac{3\mu_f}{\tilde{\mu}_p+2\mu_f}\textbf{H}_0
\end{equation}

Here $\mu_f$ represents the host medium permeability in the
ferrofluid, and the nonlinear permeability $\mu_p$ of the magnetic
particle can be expressed as $\tilde{\mu}_p=\mu_p+\xi H_1^2$ with
$\xi$ being the nonlinear coefficient. For the spherical shape
particles, the effective permeability $\mu_{f,e}$ of the whole
ferrofluid can be determined by the Maxwell-Garnett theory~\cite{18}

\begin{equation}
\frac{\mu_{f,e}-\mu_f}{\mu_{f,e}+2\mu_f}=f\frac{\tilde{\mu}_p-\mu_f}{\tilde{\mu}_p+2\mu_f}
\end{equation}

where $f$ denotes the volume fraction of the magnetic particles in
ferrofluid. Applying Eq.(2) and (3) and performing the Taylor
expansion by taking $\xi H_1^2$ as a small perturbation, we can
obtain

\begin{equation}
\mu_{f,e}=\frac{\mu_f[(1+2f)\mu_p-2(f-1)\mu_f]}{(1-f)\mu_p+(2+f)\mu_f}
+\frac{81f\mu_f^4\xi
H_0^2}{[(1-f)\mu_p+(2+f)\mu_f](\mu_p+2\mu_f)^2}+O[\xi^2]
\end{equation}

Hence the nonlinear permeability of ferrofluid can be expressed as
\begin{equation}
\tilde{\mu}=\mu_{f,e}=\mu_1+\xi_1 H_0^2
\end{equation}

in which the constants
$\mu_1=\frac{\mu_f[(1+2f)\mu_p-2(f-1)\mu_f]}{(1-f)\mu_p+(2+f)\mu_f},
\xi_1=\frac{81f\mu_f^4\xi}{[(1-f)\mu_p+(2+f)\mu_f](\mu_p+2\mu_f)^2}+O[\xi^2]$
We emphasize that when the magnetic particles form chains under
strong external field, Eq. (3) is supposed to modify by introducing
demagnetization factor~\cite{20}, but the expression of
$\tilde{\mu}$ will remain unchanged by appropriately choosing
$\mu_1$ and $\xi_1$.

\subsection{Local Magnetic Field and Induced Dipole Moment}\label{}

When the inverse ferrofluids system forms crystal under external
field, there are two dominant factors effecting its effective
permeability: the local field of crystal structure and the
geometrical shape of the particles which makes up of the crystal.
Below we will first investigate the local field factor by performing
an Ewald-Kornfeld formulation so that the structural transition and
long-range interaction can be taken into account explicitly.
Considering the ground state of inverse ferrofluids as a
body-centered-tetragonal ($bct$) lattice, which can be regarded as a
tetragonal lattice, plus a basis of two particles each of which is
fixed with an induced point magnetic dipole at its center. One of
the two particles is located at a corner and the other one at the
body center of the tetragonal unit cell, as shown in Fig.(1). Its
lattice constants are denoted by $a(=b)=\xi q^{-\frac{1}{2}}$ and
$c$= $\xi q$ along $x(y)$ and $z$ axes. In this case we identified a
different structure transformation from the $bct$ to $fcc$ to $bcc$
by changing the uniaxial lattice constant $c$ under hard-sphere
constraint. As $q$ varies the volume of unit cell remains
$v_{c}=\xi^{3}$ . In this way, the degree of anisotropy of
tetragonal lattice is measured by how $q$ is deviated from unity and
the uniaxial anisotropic axis is along the $z$ axis. In particular,
$q=0.87358$, 1 and $2^{\frac{1}{3}}$ represent the $bct$, $bcc$ and
$fcc$ lattice, respectively.

In case of an external magnetic field $H_{0}$ along the $z$ axis,
the induced dipole moment $\overrightarrow{\textbf{m}}$ are
perpendicular to the uniaxial anisotropic axis. The local field
$H_{L}$ at the lattice point $\vec{R}$=0 can be determined by using
the Ewald-Kornfeld formulation ~\cite{12,13,14},

\begin{equation}
H_{L}=\textbf{m}\cdot\sum_{j=1}^{2}\sum_{\vec{R}\neq\vec{0}}[-\gamma_{1}(R_{j})+x_{j}^{2}q^{2}\gamma_{2}(R_{j})]-\frac{4\pi
\textbf{m}}{v_{c}}\sum_{\vec{G}\neq\vec{0}}\Pi(\vec{G})\frac{G_{x}^{2}}{G^{2}}\exp\left(\frac{-G^{2}}{4\eta^{2}}\right)+\frac{4\textbf{m}\eta^{3}}{3\sqrt{\pi}}
\end{equation}
In this equation, $\gamma_{1}$ and  $\gamma_{2}$ are two
coefficients, given by,
\begin{equation}
\gamma_{1}(r)=\frac{\rm erfc(\eta
r)}{r^{3}}+\frac{2\eta}{\sqrt{\pi}r^{2}}\exp(-\eta^{2}r^{2})
\end{equation}

\begin{equation}
\gamma_{2}(r)=\frac{3\rm erfc( \eta
r)}{r^{5}}+(\frac{4\eta^{3}}{\sqrt{\pi}r^{2}}+\frac{6\eta}{\sqrt{\pi}r^{4}})\exp(-\eta^{2}r^{2})
\end{equation}
where erfc$(\eta r)$ is the complementary error function, and $\eta$
is an adjustable parameter making the summations in real and
reciprocal lattices  converge rapidly. In Eq. (7), $R$ and $G$
denote the lattice vector and reciprocal lattice vector,
respectively,

\begin{equation}
\vec{R}=\xi(q^{-\frac{1}{2}}l\hat{x}+q^{-\frac{1}{2}}m\hat{y}+qn\hat{z})
\end{equation}

\begin{equation}
\vec{G}=\frac{2
\pi}{\xi}(q^{\frac{1}{2}}u\hat{x}+q^{\frac{1}{2}}v\hat{y}+q^{-1}w\hat{z})
\end{equation}
where, $l, m, n, u, v, w$ are integers. In addition, $z_{j}$ and
$R_{j}$ are respectively given by,

\begin{equation}
z_{j}=l-\frac{j-1}{2},
R_{j}=|\vec{R}-\frac{j-1}{2}(a\hat{x}+a\hat{y}+c\hat{z})|
\end{equation}
and $\Pi(\vec{G})=1+\exp[i(u+v+w)/\pi]$ is the structure factor. The
local field will be computed by summing over all integer indices
$(j, l, m, n) \neq (1, 0, 0, 0)$ for the summation in real lattice
and $(u, v, w) \neq (0, 0, 0)$ for the summation in the reciprocal
lattice. Here we consider an infinite lattice. For finite lattices,
one must be careful about the effects of different boundary
conditions ~\cite{15}. Because of the exponential factors, we may
impose an upper limit to the indices, i.e., all induces ranging from
$-L$ to $L$, where $L$ is a positive integer.

Now, let us define a local field factor $\alpha$,

\begin{equation}
\alpha=\frac{3v_{c}}{4\pi}\frac{H_{L}}{\textbf{m}}
\end{equation}

It is worth noticing that $\alpha$ is a function of single variable
$q$. Also, there is a sum rule 2$\alpha_{\perp}$
+$\alpha_{\parallel}$ =3 ~\cite{16,17}, where $\alpha_{\parallel}$
denotes the local field factor in longitudinal field case, while
$\alpha_{\perp}$ denotes the local field in transverse case. Here
the longitudinal (or transverse) field case corresponds to the fact
the $H_{0}$ field is parallel (or perpendicular) to the uniaxial
anisotropic $z$ axis.

\subsection{Nonlinear AC response in the Cases of Spheroids}\label{}

Now we are in the position to take into the account the geometrical
shape effect of the particles by investigating nonspheres: prolate
and oblate. In MR(ER or inverse ferrofluids) fluid, the shear stress
and permeability(permittivity) will be enhanced by changing the
ratio of the ellipsoidal particles in the chain ~\cite{Shaw} such as
$p$-phenylene-2, 6- benzobisthiazole polyparticle. This induced
polarization can be described in demagnetizing factor, which
includes the three principal axes of the spheroid. For a prolate
spheroid($a=b<c$), the longitudinal demagnetizing factor $g_l$ along
the major axis $c$ is given by $g_l=\frac{k \rm
ln(k+\sqrt{k^2-1})}{(k^2-1)^{\frac{3}{2}}}-\frac{1}{k^2-1}$ with
$k=\frac{c}{a}$; for oblate case($a<b=c$), $g_l=1-\frac{k^2\rm
arcsin\frac{\sqrt{k^2-1}}{k}}{(k^2-1)^{\frac{3}{2}}}+\frac{1}{k^2-1}$.
There is a sum rule for the longitude and transverse demagnetizing
factor~\cite{18,20} that $g_l+2g_t=1$. For prolate spheroid, because
it can easily align with its major axis($c$) along the direction of
the external field, while the dipole moments of nonmagnetic
particles are oriented opposing the external field, here we only
focus on the longitude case.  It is also worth noticing that the
spherical shape of particles is included when $g_l=g_t=\frac{1}{3}$.

Now We can use the developed Maxwell-Garnett theory for uniaxially
anisotropic suspensions ~\cite{18,19} considering both the local
lattice and geometrical shape effect to derive the effective
permeability~\cite{20}

\begin{equation}
\frac{\frac{1}{3}\alpha(\mu_e-\tilde{\mu})}{\tilde{\mu}+\frac{1}{3}\alpha(\mu_e-\tilde{\mu})}
=p\frac{g_l(\mu_0-\tilde{\mu})}{\tilde{\mu}+g_l(\mu_0-\tilde{\mu})}
\end{equation}

Similar to the technique applying in Eq.(4), combining Eq.(1) we
admit

\begin{equation}
\mu_e=\frac{\mu_1[\alpha\mu_1+g_l(x\alpha-3p)y]}{\alpha[(\mu_1+g_lxy]}+\xi
H_0^2\frac{\alpha\mu_1^2+g_l^2xy^2[x\alpha-3p]+2\alpha g_l\mu_1
xy}{\alpha[yg_l-\mu_1]^2[\mu_1+g_lxy]^2}+\frac{3g_l^2\mu_0^2\mu_1^4px
\xi^2H_0^4}{\alpha[\mu_1-g_ly]^4[\mu_1+g_lxy]^3} +O[\xi^3]
\end{equation}

with $x=p-1,y=\mu_1-\mu_0$.

In the presence of external oscillating time-varying magnetic
field~cite\cite{21}, the magnetic particles will have nonlinear
characteristics and the inverse ferrofluids system will induce
harmonic response at different frequencies without phase
transition~\cite{10}. In the experiment, the second term or higher
order terms on the right-hand of Eq. (14) can be obtained using
$\emph{mixed-frequency measurements}$~\cite{yang}. In this paper, we
will focus on the characteristics of the harmonics of effective
permeability for the inverse ferrofluid and their nature on the
magnetophoresis when particles put in: $\mu_{0}, \mu_{\omega},
\mu_{2\omega}...$. In case of an external sinusoidal magnetic
field(ac) of the form

\begin{equation}
H_0=H_{ac}\rm sin\omega t
\end{equation}

Applying Eq.(15) into Eq.(14), we can obtain zero, second and fourth
order harmonics for the effective permeability:
$\mu_e=\mu_0+\mu_{2\omega}\rm cos2\omega t+\mu_{4\omega}\rm
cos4\omega t$. Here we give the expression for $\mu_{4\omega}$:

\begin{equation}
\mu_{4\omega}=\xi
H_{ac}^2\frac{3g_l^2\mu^4px}{8\alpha[g_lxy+\mu]^3[\mu-g_ly]^4}
\end{equation}

On the other hand, under the application of a dc and ac field,
namely,

\begin{equation}
H_0=H_{dc}+H_{ac}\rm sin\omega t
\end{equation}

Using the same technique we can rewrite the effective permeability
as $\mu_e=\mu_0+\mu_{\omega}\rm sin\omega t+\mu_{2\omega}\rm
cos2\omega t+\mu_{3\omega}\rm sin3\omega t+\mu_{4\omega}\rm
cos4\omega t$, for extracting different harmonics.

\subsection{Magnetophoresis in inverse ferrofluids system}\label{}

Now we will investigate the magnetophoretic behavior when particles
are put in inverse ferrofluids. In a standard case, a magnetically
polarizable object will be trapped in a region of a focused magnetic
field, or nonmagnetic particles in a magnetizable
liquid(ferrofluid), provided there is sufficient magnetic response
to overcome thermal energy and the magnetophoretic force~\cite{1}.
Here we consider the nonmagnetically homogenous particle in the
nonlinear inverse ferrofluids for simplicity, and the magnetic field
is assumed to be $dc$ and hysteresis is ignored. For a magnetically
linear particle under magnetophoresis, the effective magnetic dipole
moment vector induced inside takes a form very similar to that for
the effective dipole moment of dielectric paricle~\cite{2,chou},
$\vec{m}=4\pi R_{eff}^{3}b\vec{H}_{0}$, where $R_{eff}$ is the
effective radius of spherical or spheroidal particle and $b$ is the
clausius-Mossotti factor(CMF) along the direction of external field.
Here we focus on the magnetophoresis of spherical particle in
inverse ferrofluids, and CMF $b$ is written as:
$b=\frac{\mu_2-\mu_e}{\mu_2+2\mu_e}$ where $\mu_{2}$ is the
permeability of nonmagnetic particle, and $\mu_{e}$ is the
permeability of the host medium: inverse ferrofluids. It should be
noted, that in a general case, for spheroidal particle, CMF $b$ is
written as $\frac{1}{3}\frac{\mu_2-\mu_e}{\mu_e+g_l(\mu_2-\mu_e)}$.
The force exerted by a nonuniform magnetic field on the dipole can
be obtained:

\begin{equation}
\vec{F}_{{dipole}}=\mu_{e}\vec{m}\cdot\nabla H_{0}
\end{equation}

Combining the effective magnetic dipole moment, the magnetophoretic
force exerted on the spherical particle in a nonuniform magnetic
field can be written as,

\begin{equation}
\vec{F}_{MAP}=2\pi\mu_{e}R^{3}b\nabla H_{0}^{2}
\end{equation}

Thus the magnetophoretic force on particles are proportional to the
particle volume, the polarization difference and distribution of the
geometric field gradient. There are two kinds of magnetophoresis
depended upon the relative magnitudes of $\mu_{2}$ and $\mu_{e}$.
Particles are attracted to magnetic field intensity maxima and
repelled to the field-generating electrodes when $b>0$ as positive
magnetophoresis and negative magnetophoresis corresponds to $b<0$.
Based on the harmonics of effective permeability, the expressions
for the CMF can explicitly be obtained when the lattice structure of
the host medium is changed, as shown in Fig. 2. The degree of
anisotropy $q$ plays an important role in determining the local
field factor $\alpha$. There is a plateau during the structure
transformation on the relationship between $q$ and
$\alpha$~\cite{Huang}, accordingly we can also predict similar
behavior in CMF, especially in Fig. 2(b). It is found that
increasing $q$ causes both zero and fourth order harmonic CMF
decreases in Fig. 2(a) and Fig. 2(c). Note that when $q$ changes
from 0.6 to 0.8 all CMF have great decreasing and CMF in Fig. 2(c)
changes its sign when the nonlinearity is the highest, thus
demonstrating the sensitivity of CMF when the system is under
structure changes. Although the fourth-order harmonics of the local
magnetic field and induced dipole moment seems to be complicated,
since the strength of the nonlinear polarization of nonlinear
materials can be reflected in the magnitude of the high-order
harmonics of local magnetic field and dipole moment, the effective
nonlinear part in the effective nonlinear magnetic constant can be
expected to determine the magnitude of the fourth-order harmonics.

\section{Numerical Results}\label{}

In the preceding section we will calculate the effect of spheroidal
shape and volume fraction of nonmagnetic particles which forms bct
lattice structure in the ferrofluids, on the harmonic response of
CMF. For a special case, when we would like to conclude the
nonharmonic response situation, the external field $H_0$ is constant
and not vary with time namely, it can be easily obtained without
tedious calculation by setting $\omega t=\frac{\pi}{2}$ for ac field
in Eq. (15). Thus only the zero harmonic contributes. Similar
conclusions can be obtained that only $b_0+b_{\omega}-b_{3\omega}$
contributes in a dc and ac field case.

In Fig. 3 and Fig. 4 we show the harmonic response of CMF for
different shapes in ac field: prolate and oblate. The zero(or
nonharmonic) response of CMF exhibits strong sensitivity to the
prolate demagnetization factor(the shape) and some peaks are
obtained, while for oblate CMF $b_0$ have slight change
compared(Fig. 4). This can be briefly explained through
self-consistent approach~\cite{22} by the spectral representation
separating the material parameter $s=(1-\mu_0/\mu)$ from the
geometrical parameters. Without considering the crystal lattice
effect, the effective permeability can be derived as
$\mu_e=\mu[1-\frac{p}{s-g_l(1-p)}]$. It is clear that decreasing
$g_l$ leads to increasing $\mu_e$, while for prolate(oblate) case,
increasing aspect ratio $k$ causes increasing(decreasing) of $g_l$.
In the meantime, because of the existence of weak nonlinearity, the
nonlinear polarization has just a perturbation effect, which can be
neglected when compared to the linear part. Furthermore, we can
predict that one or two shape-dominant crossover ratio at which
there is no net force on the particle in the magnetophoresis. In
Fig. 3(a), it is interesting that the crossover ratio is
monotonically increasing as the nonlinearity increases. Fig. 4 shows
that, as nonlinear characteristics increase, the harmonics response
of CMF are caused to increase accordingly for oblate particles. We
emphasize to point out that, although prolate and oblate particles
can both exist in different system state, the alignment of one
particle can only be along the longest axis(prolate) when the
particle is \emph{homogenous} under the external magnetic field.

Fig. 5 and Fig. 6(a)-(e) show different harmonic response of CMF for
prolate and oblate particle in dc and ac field. Comparing with in
single ac field, the response nature of CMF keeps unchanged for the
same harmonics. So the introduction of coupling between dc and ac
field add more odd harmonic response. As discussed above the
nonharmonic response of CMF can be obtained as
$b_0+b_{\omega}-b_{3\omega}$, which is shown in Fig. 5(f) and Fig.
6(f).It is also found that lower harmonic response of CMF can be
several orders of magnitude larger than the higher one.

Fig. 7 displays for prolate, oblate spheroidal and spherical
particle, the effect of volume fraction of magnetic particles in the
ferrofluids on the harmonic response of CMF. It is observed from
Fig. 7(a) that the nonlinearity for nonharmonic response has little
effect on CMF. The reason is that in the system of interest only
ferrofluids are assumed to be nonlinear while the particles are
linear. As the volume fraction of particles increases, the nonlinear
component within the system have slight change naturally, yielding
unchanged CMF accordingly. The second and fourth order response of
CMF in Fig. 7(b) and (c) shows great variety when the volume
fraction changes. This enhancement of the harmonics may arise from
the mutual interaction between the dipole moments: when increasing
the volume fraction, the mutual interaction becomes strong and leads
to shift of harmonic response. It is observed that for prolate
particles, the nature of zero-order response of CMF is greatly
different from that of oblate spheroidal and spherical particles,
comparing Fig. 7(a) with Fig. 7(d) and Fig. 7(g).

\section{Discussion and Conclusion}\label{}

In this work, we have investigated both the effect of lattice
structure of crystal and the geometrical shape of particles embedded
in the ferrofluids on the nonlinear response of magnetophoresis
using the perturbation method. The nonlinear response of CMF under
the structure transition is studied in detail. It is thus possible
to perform a real-time monitoring of structure transformation. We
find that the coupling of ac and dc field case can lead to
fundamental and third harmonic response in the effective
permeability constant and the CMF respectively. We also find that
change the aspect ratio in both prolate and oblate particles can
alter the harmonic and nonharmonic response and thereby causing the
magnetophoretic force vanish. Furthermore, by taking into account
the local-field effect from the mutual interaction, we examine the
CMF response in various volume fractions, and find that increasing
volume fraction can result in the enhancement of the strengths and
the difference between the different nonlinearities and
characteristic harmonics. Thus, it is possible to obtain a good
agreement between theoretical predictions and experimental data by
suitable adjustment of both the geometric parameters(for example,
the particle shape) and the physical parameters(for example, the
temperature, component ratio in the system, saturation
magnetization, initial magnetic susceptibility). On the contrary,
such fitting is useful to obtain the relevant physical information
of magnetophoresis in inverse ferrofluid.

We demonstate theoretically the shape effect of the effective
magnetic constant in the inverse ferrofluids system. It is in fact
that the magnetic particles in the ferrofluid are inhomogeneous in
the real case. A first-principle study of the dielectric dispersion
of inhomogeneous suspension is presented~\cite{Huang,norina}, and
can be used to analyze the magnetic dispersion in the dilute limit,
in which the permeability of the ferrofluid is dependent on the
frequencies of external magnetic field. Also when the inhomogeneity
is simplified as the shelled spheroidal particles in the ferrofluid
system~\cite{Bizdoaca}, a self-consistent method~\cite{23} can be
applied for the three-component composites. The magnetic anisotropy
as well as geometric anisotropy can both be taken into account, and
will be reported elsewhere.

The nonlinearity under our consideration is weak, which is common in
real situations under moderate fields, while in case of strong
nonlinearity, the perturbation approach is no longer valid and
should be adopted using self-consistent method instead~\cite{24}.
And in the above discussion, we consider the longitude case under
the external field, it can also be found constricted by the relation
$g_l+2g_t=1$, the harmonic response of CMF have opposite behavior.

\section*{Acknowledgements}

The author Y.C.J is grateful to Prof. TM. Hong for his generous help
and hospitality at NTHU in Taiwan in the academic year 2006
supported by ChunTsung(T. D. Lee) Foundation and would like to
express the gratitude to Prof. Chia-Fu Chou for the stimulating
discussions in Wu-ta you Camp and Sinica. We would like to thank
Prof. Montgomery Shaw and R. Tao for great support. The authors
acknowledge the financial support by the Shanghai Education
Committee and the Shanghai Education Development Foundation (¡±Shu
Guang¡± project) under Grant No. KBH1512203, by the Scientific
Research Foundation for the Returned Overseas Chinese Scholars,
State Education Ministry, China, by the National Natural Science
Foundation of China under Grant No. 10321003.

\newpage
\section*{Figure Captions}

Fig.1. (Color online) Schematic graph showing the inverse ferrofluid
system, in which the crystal consisted of nonmagnetic
particles(permeability is denoted by $\mu_0$) embed in the
ferrofluid($\tilde{\mu}_0$). Ferrofluids contain nanosize
ferromagnetic particles($\tilde{\mu}_p$) dispersed in a carrier
fluid($\mu_f$).

Fig.2. (Color online) The zero, second, fourth harmonic response of
CMF versus degree of anisotropy $q$ for different intrinsic
nonlinear characteristics in the prolate case. The bct, bcc and fcc
lattices which are respectively related to $q=0.87358, 1.0,
2^{\frac{1}{3}}$ are shown(dot lines). Parameters: $\mu_0=\mu_2=1,
\mu_1=2, p=0.1, k=1.2$.

Fig.3. (Color online) Cases of prolate spheroidal particles: ac
field. Zero, second, fourth harmonic response of CMF versus $k=c/a$
for different intrinsic nonlinear characteristics. Parameters:
$\mu_0=\mu_2=1, \mu_1=2, p=0.1, \alpha=0.95351$.

Fig.4. (Color online) Cases of oblate spheroidal particles: ac
field. Zero, second, fourth harmonic response of CMF versus $k=c/a$
for different intrinsic nonlinear characteristics. Parameters:
$\mu_0=\mu_2=1, \mu_1=2, p=0.1, \alpha=0.95351$.

Fig.5. (Color online) Cases of prolate spheroidal particles: ac and
dc field. Zero, fundamental, second, third, fourth harmonic response
of CMF versus $k=c/a$ for different intrinsic nonlinear
characteristics (a)-(e). Parameters: $\mu_0=\mu_2=1, \mu_1=2, p=0.1,
\alpha=0.1, H_{dc}=\sqrt{3.6}$. (f) show the nonharmonic response of
CMF.

Fig.6. (Color online) Cases of oblate spheroidal particles: ac and
dc field. Zero, fundamental, second, third, fourth harmonic response
of CMF versus $k=c/a$ for different intrinsic nonlinear
characteristics (a)-(e). Parameters: $\mu_0=\mu_2=1, \mu_1=2, p=0.1,
\alpha=0.95351, H_{dc}=\sqrt{3.6}$. (f) show the nonharmonic
response of CMF.

Fig.7. (Color online) Zero, second, fourth harmonic response of CMF
versus volume fraction $p$ for different intrinsic nonlinear
characteristics in cases of prolate((a)-(c)), oblate((g)-(i))
spheroidal particles and spherical particles((d)-(f)) under ac
field. Parameters: $\mu_0=\mu_2=1, \mu_1=2, k=1.2, \alpha=0.95351$.

\newpage
\begin{figure}[h]
\includegraphics[width=400pt]{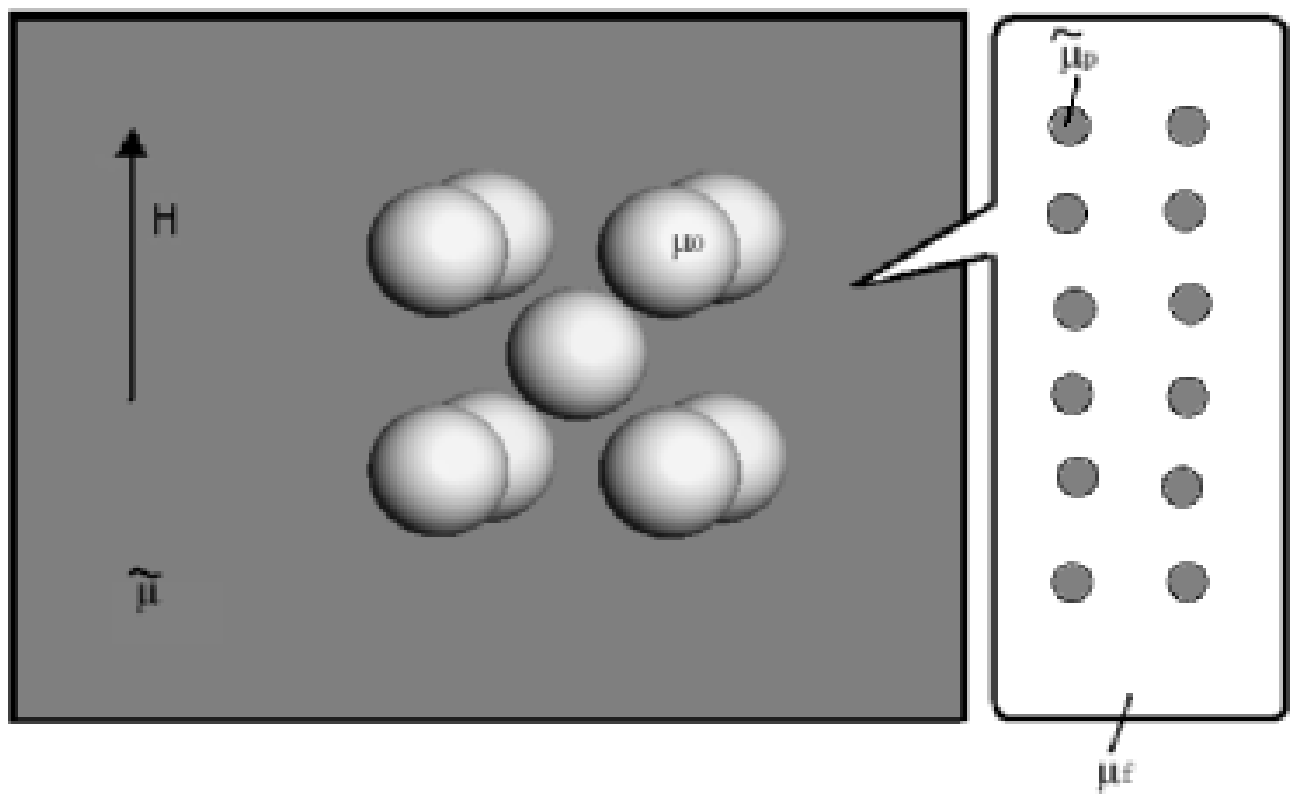}
\caption{Jian, Gao, and Huang}.\label{}
\end{figure}

\newpage
\begin{figure}[h]
\includegraphics[width=300pt]{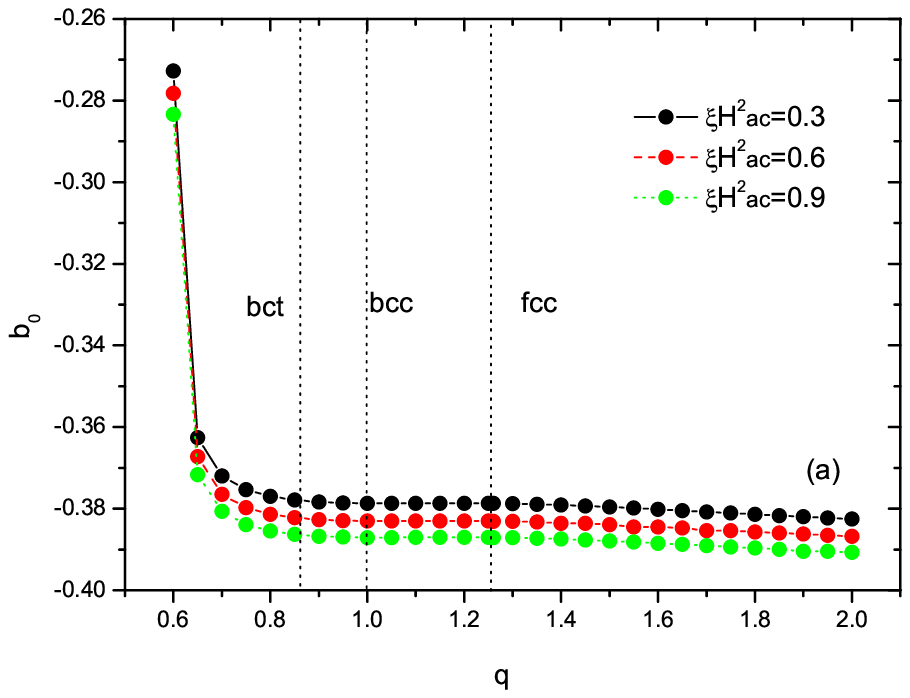}
\includegraphics[width=300pt]{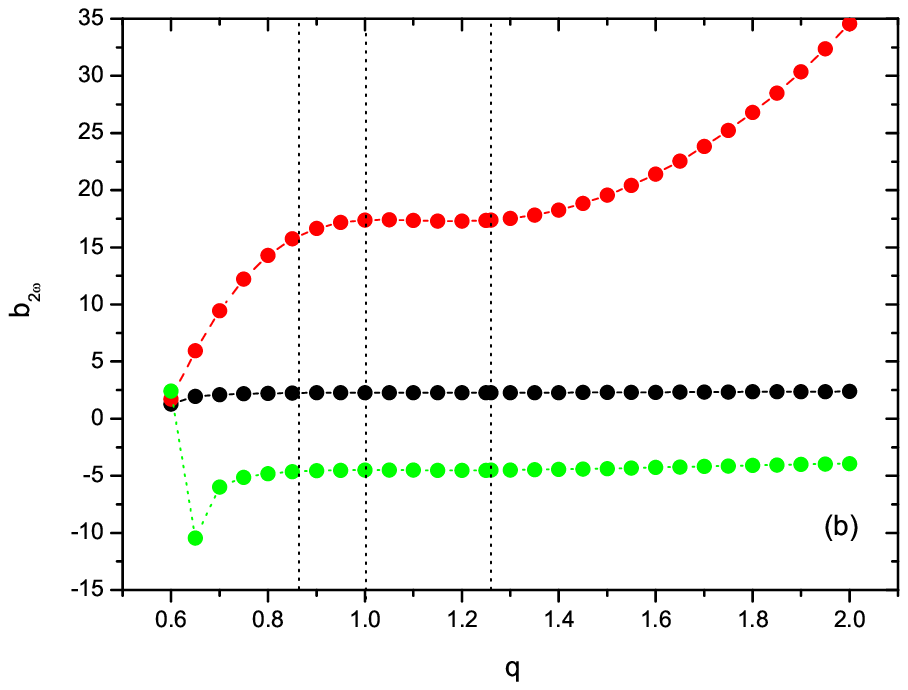}
\includegraphics[width=300pt]{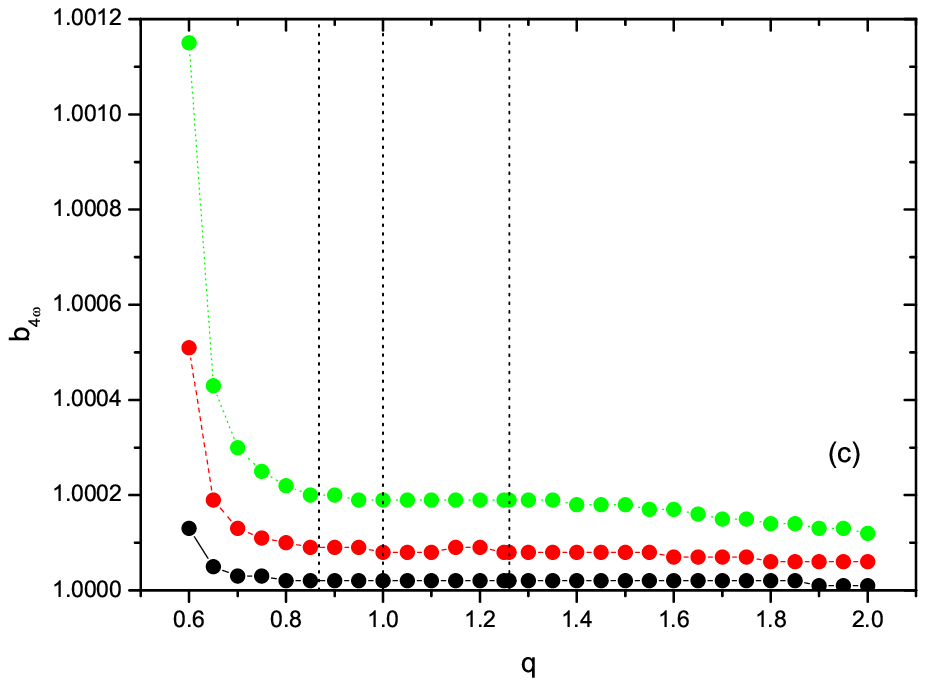}
\caption{Jian, Gao, and Huang}\label{}
\end{figure}

\newpage
\begin{figure}[h]
\includegraphics[width=300pt]{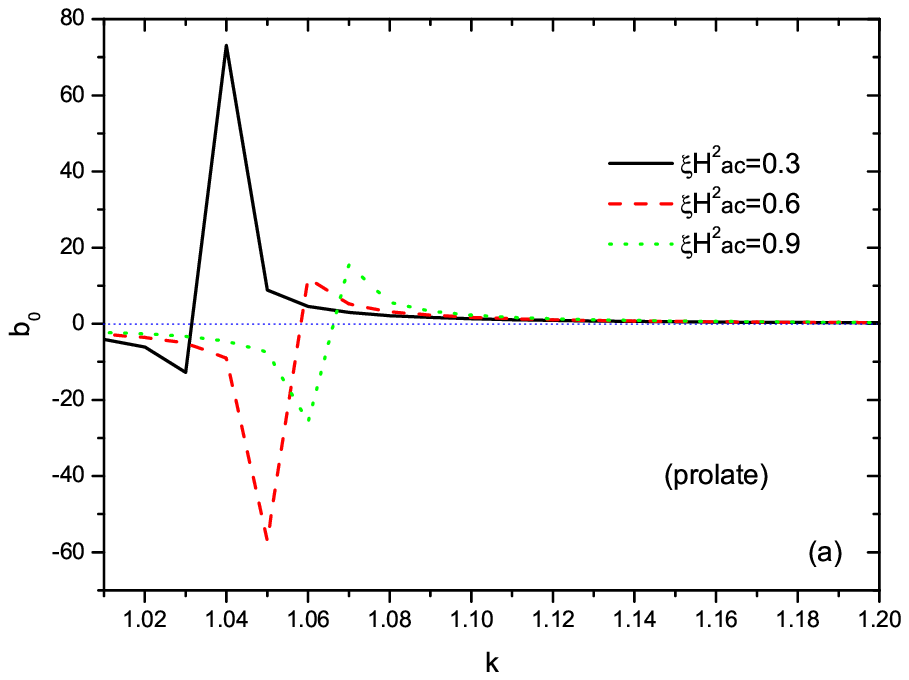}
\includegraphics[width=300pt]{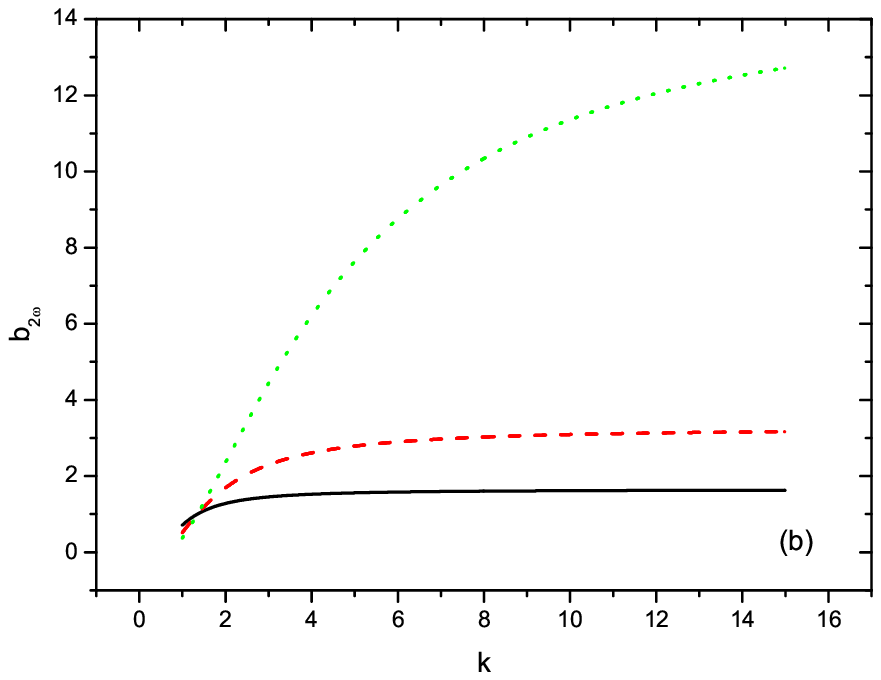}
\includegraphics[width=300pt]{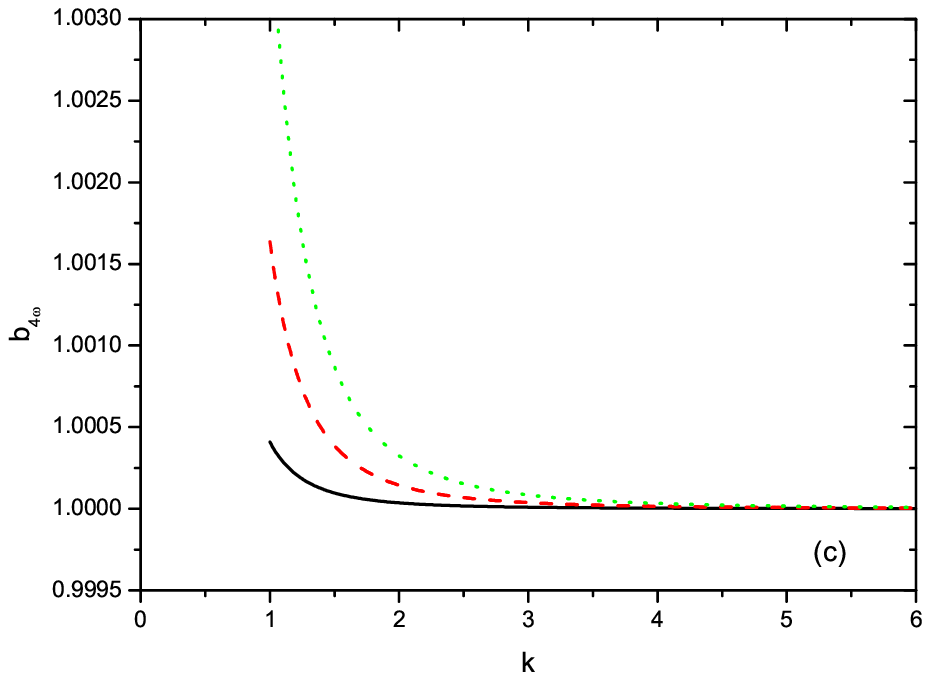}
\caption{Jian, Gao, and Huang}\label{}
\end{figure}

\begin{figure}[h]
\includegraphics[width=300pt]{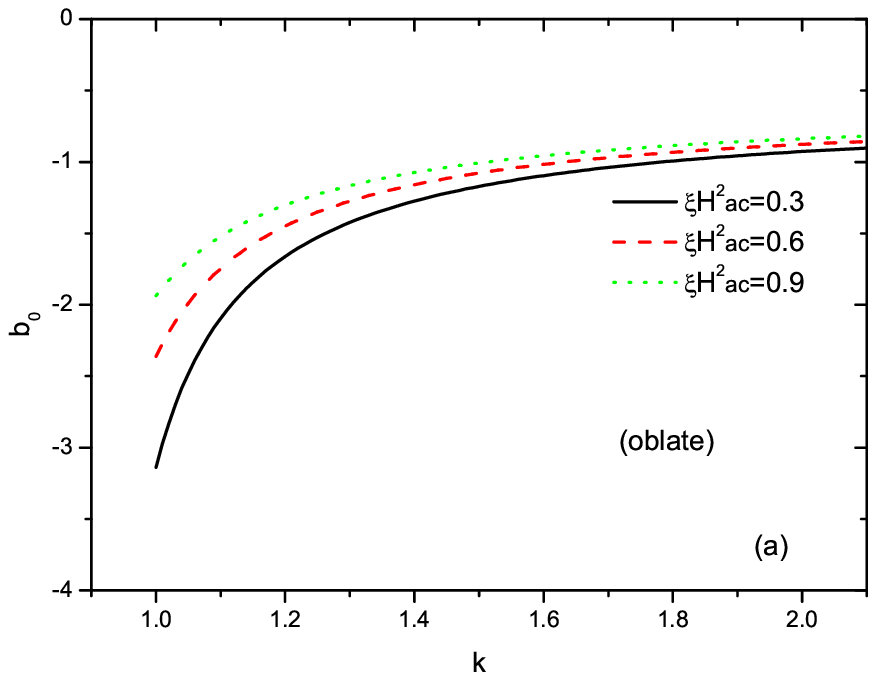}
\includegraphics[width=300pt]{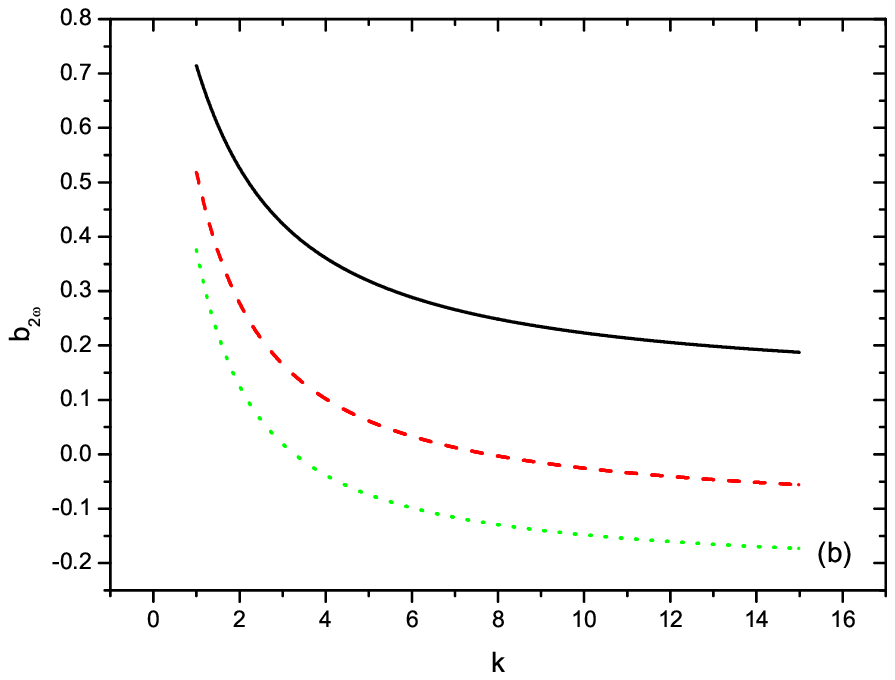}
\includegraphics[width=300pt]{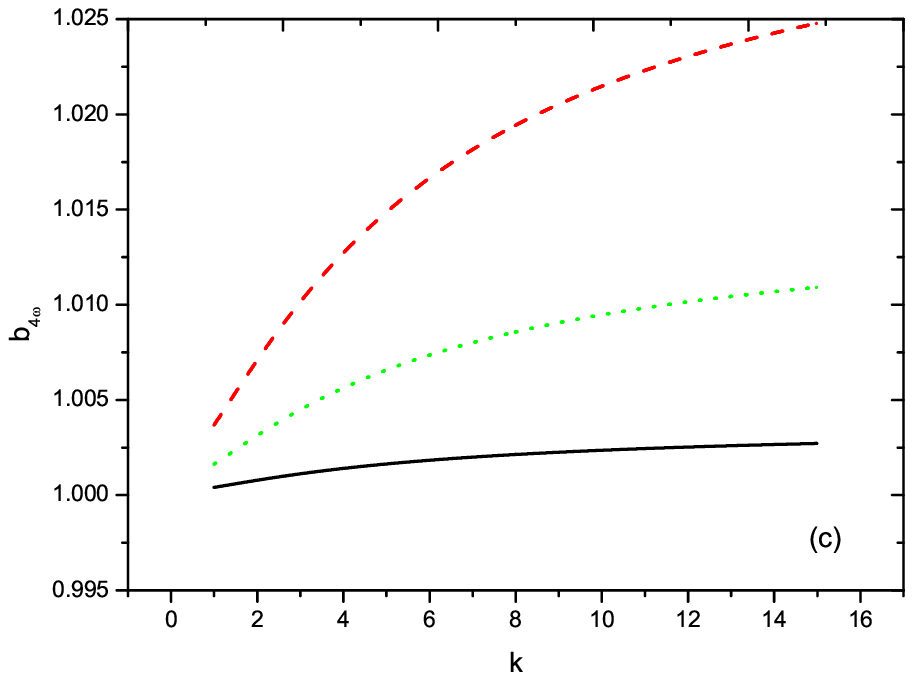}
\caption{Jian, Gao, and Huang}\label{}
\end{figure}

\begin{figure}[h]
\includegraphics[width=200pt]{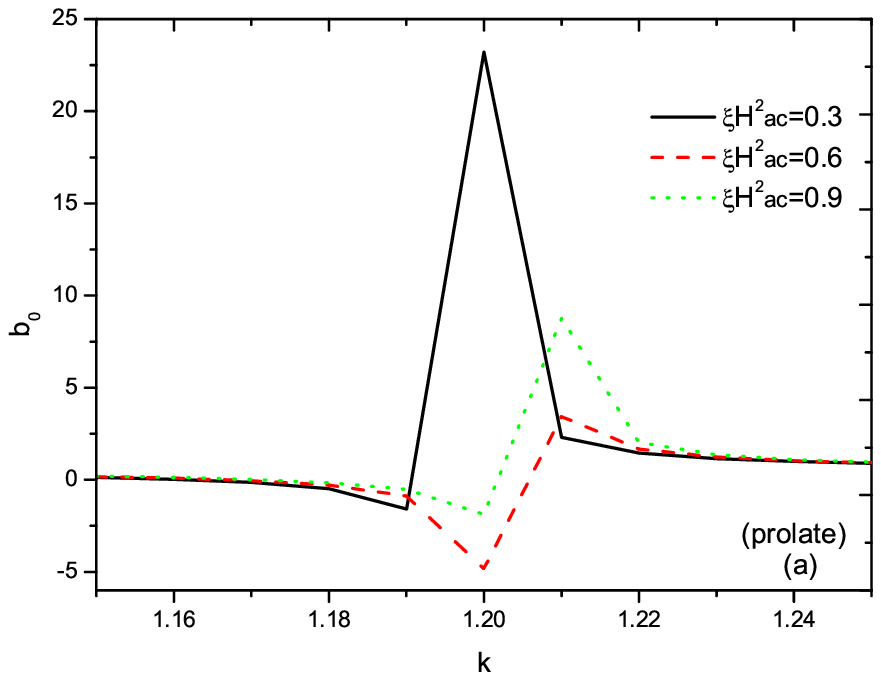}
\includegraphics[width=200pt]{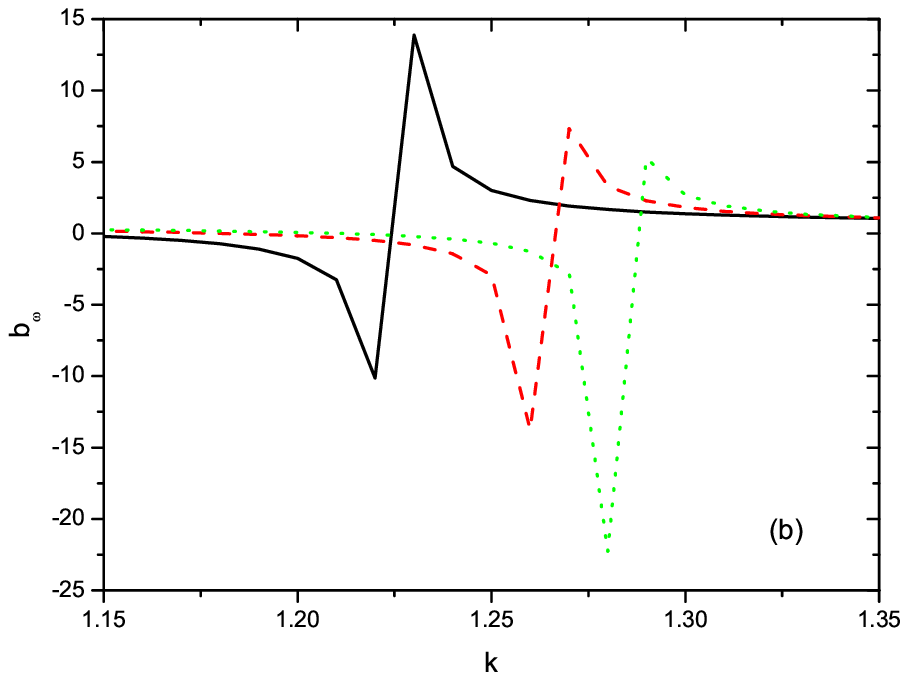}
\includegraphics[width=200pt]{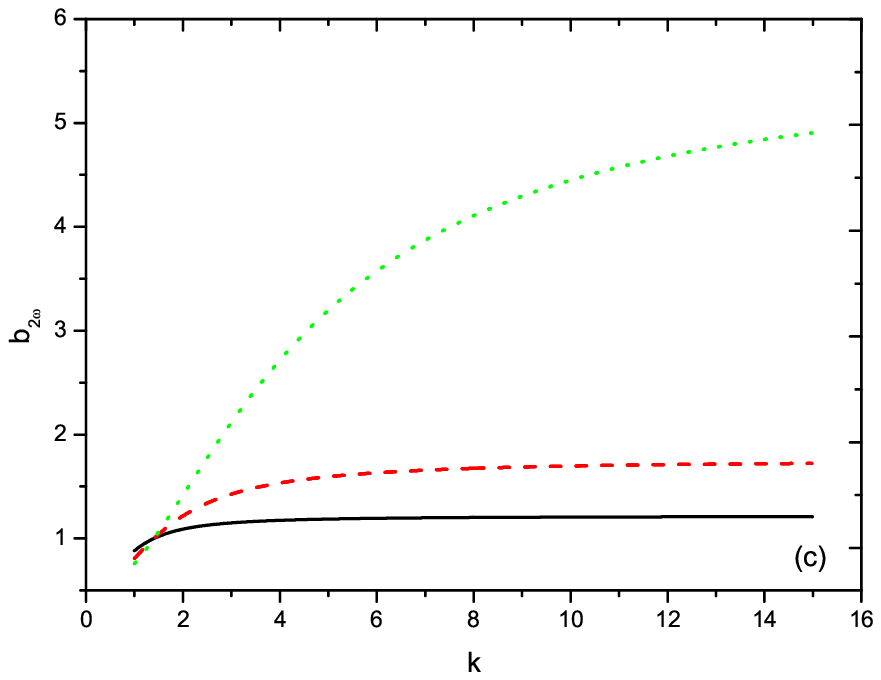}
\includegraphics[width=200pt]{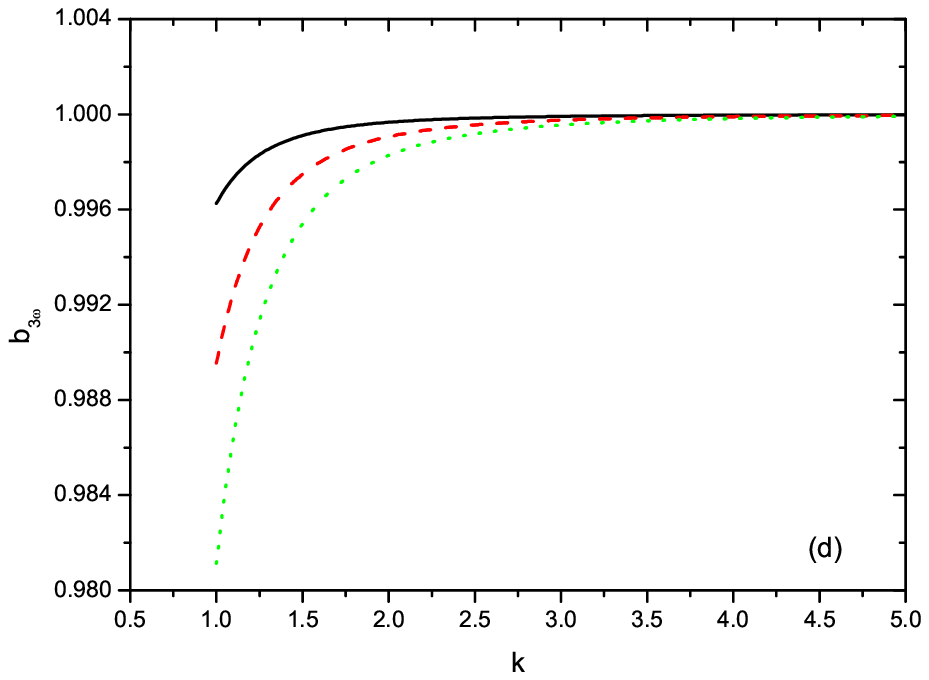}
\includegraphics[width=200pt]{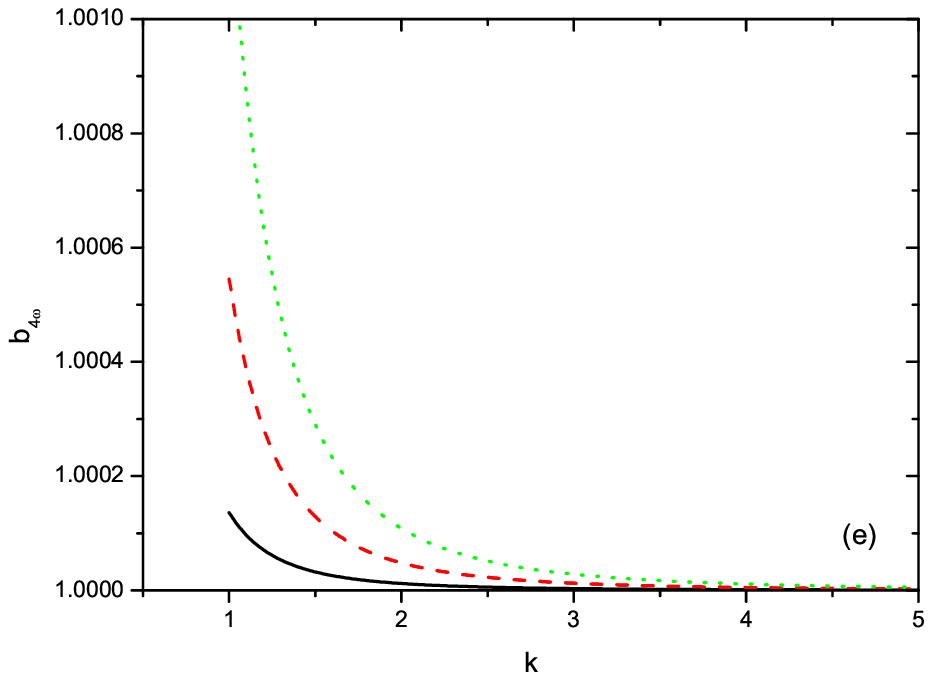}
\includegraphics[width=200pt]{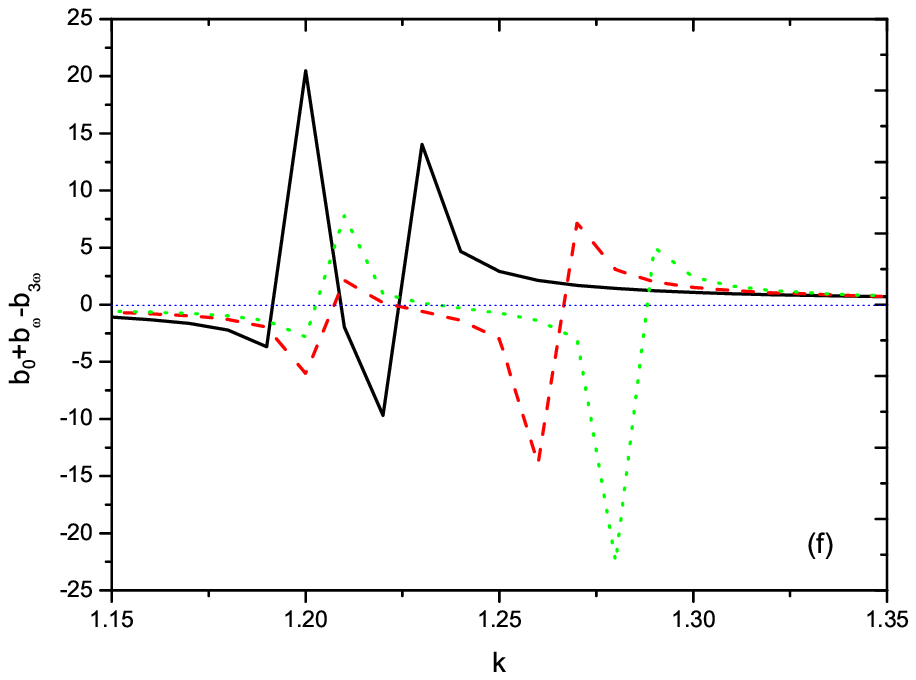}
\caption{Jian, Gao, and Huang}\label{}
\end{figure}

\newpage
\begin{figure}[h]
\includegraphics[width=200pt]{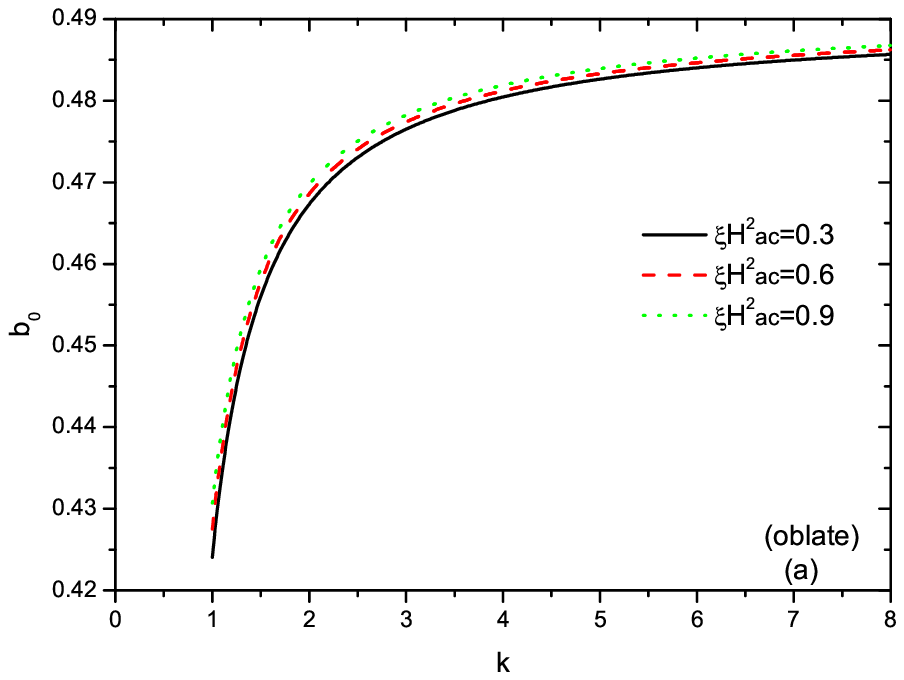}
\includegraphics[width=200pt]{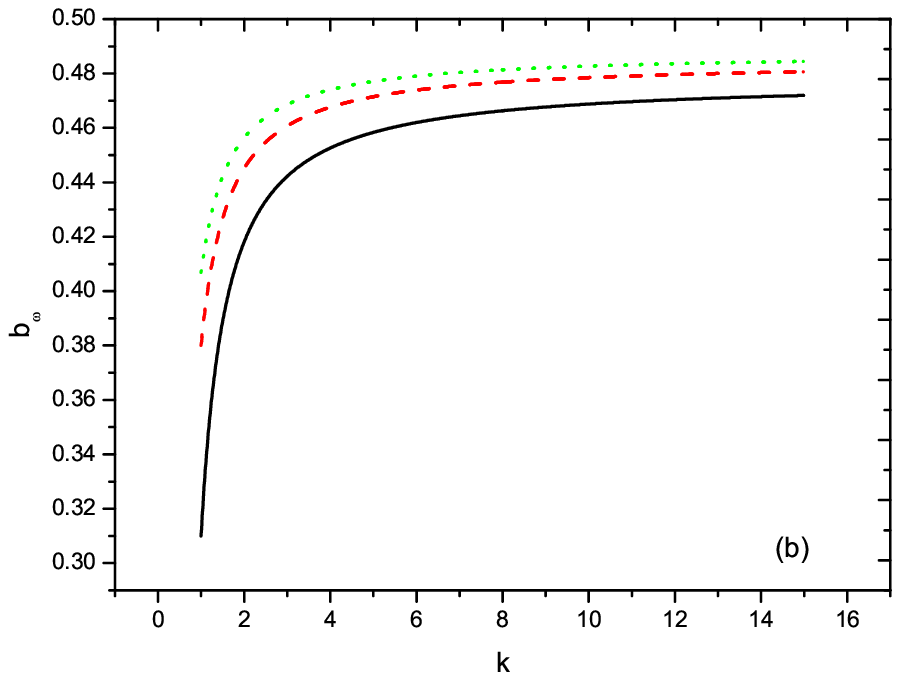}
\includegraphics[width=200pt]{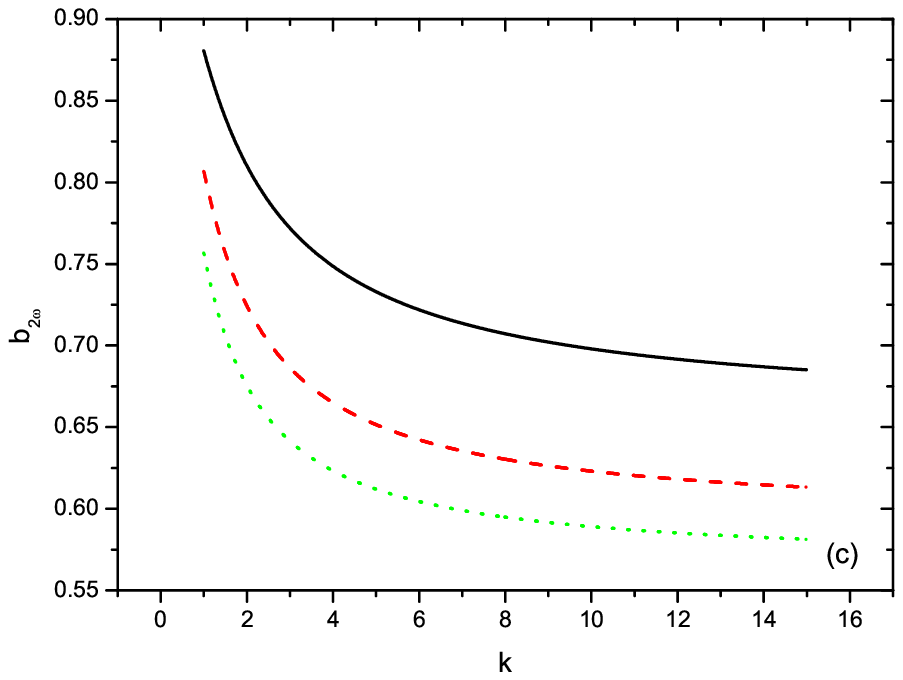}
\includegraphics[width=200pt]{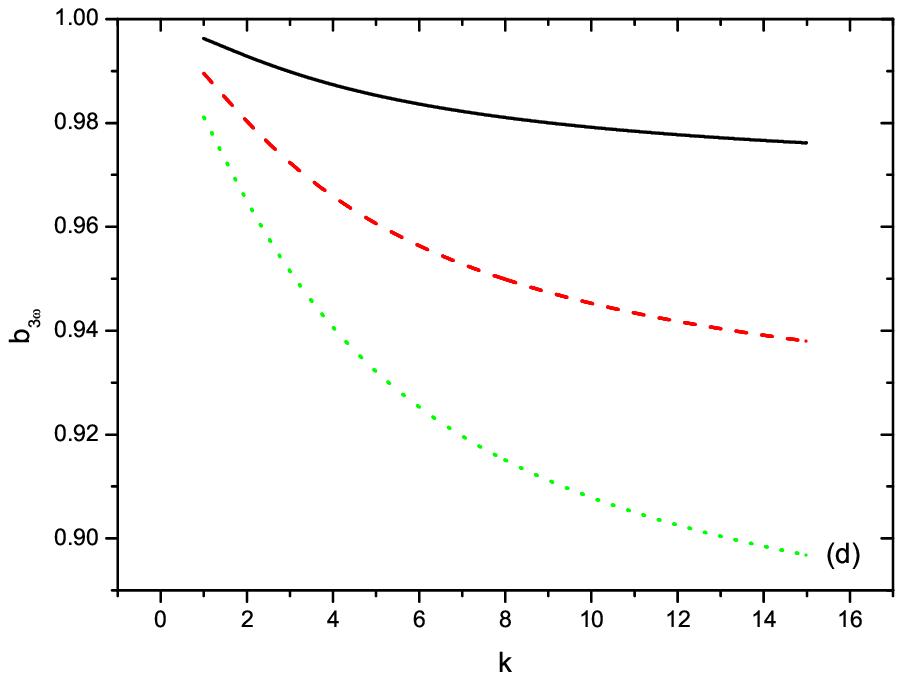}
\includegraphics[width=200pt]{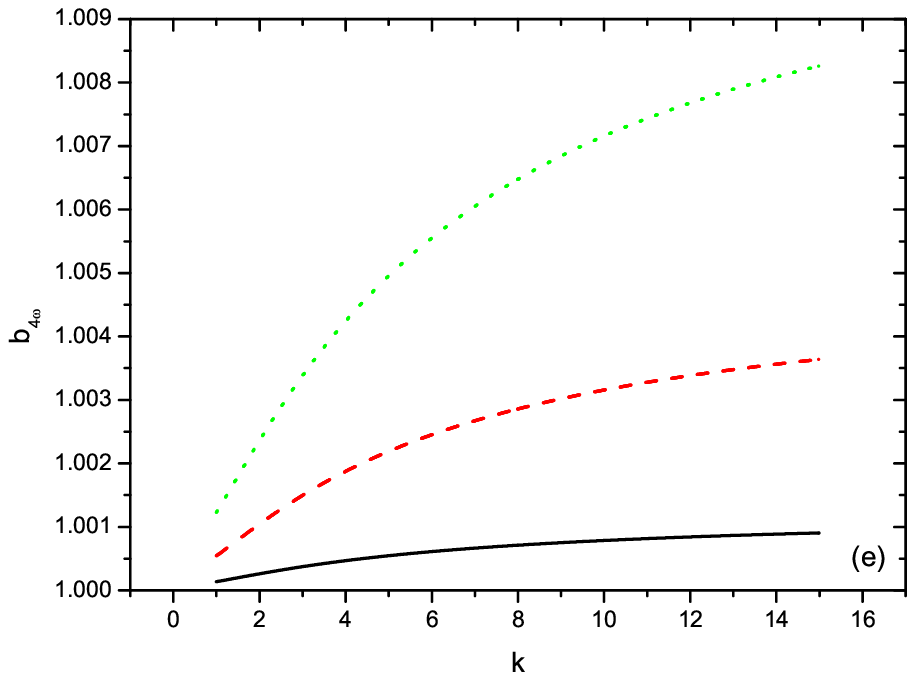}
\includegraphics[width=200pt]{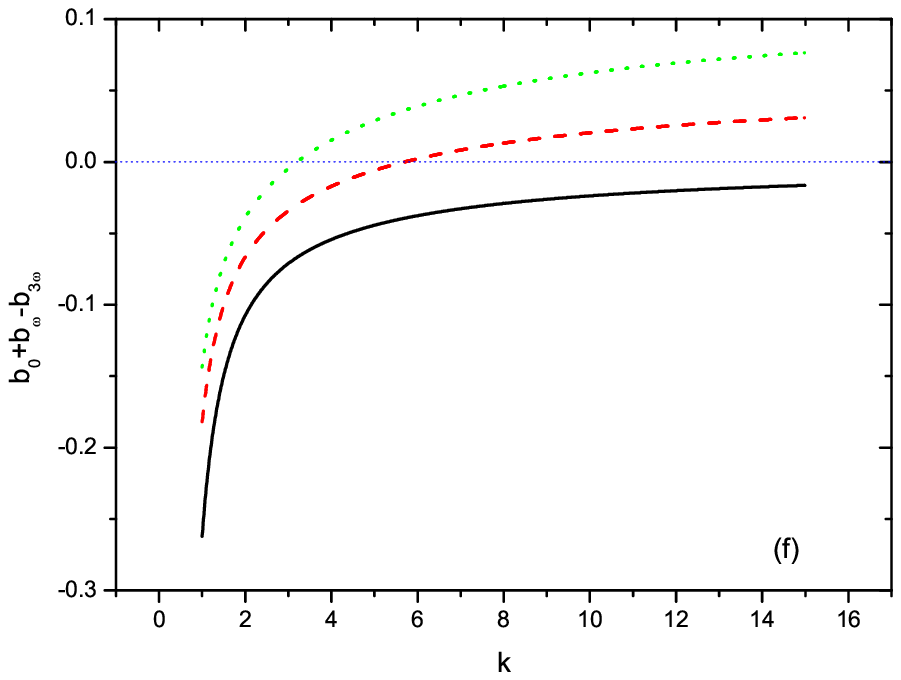}
\caption{Jian, Gao, and Huang}\label{}
\end{figure}

\newpage
\begin{figure}[h]
\includegraphics[width=150pt]{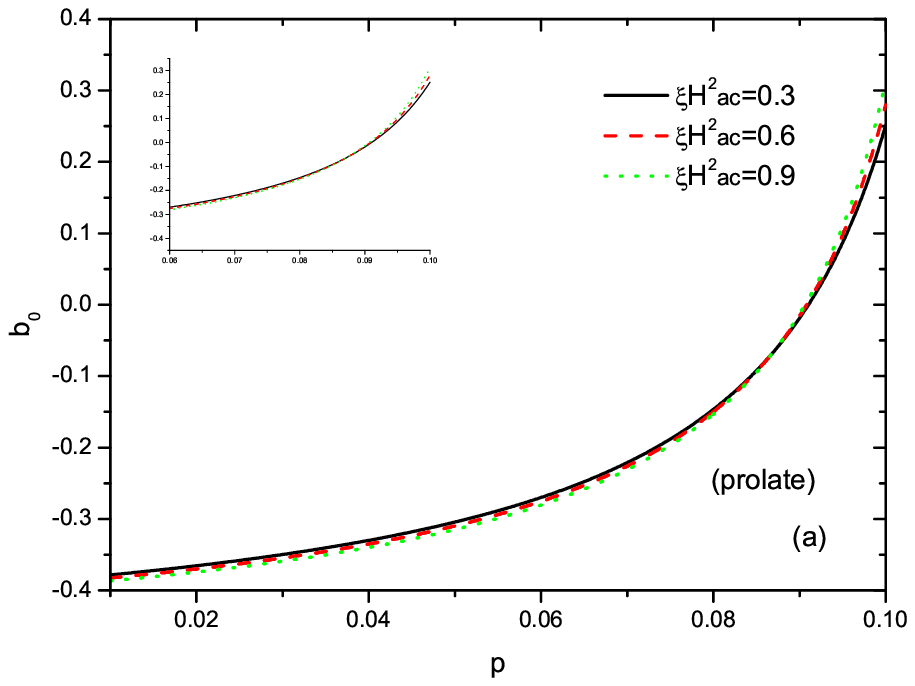}
\includegraphics[width=150pt]{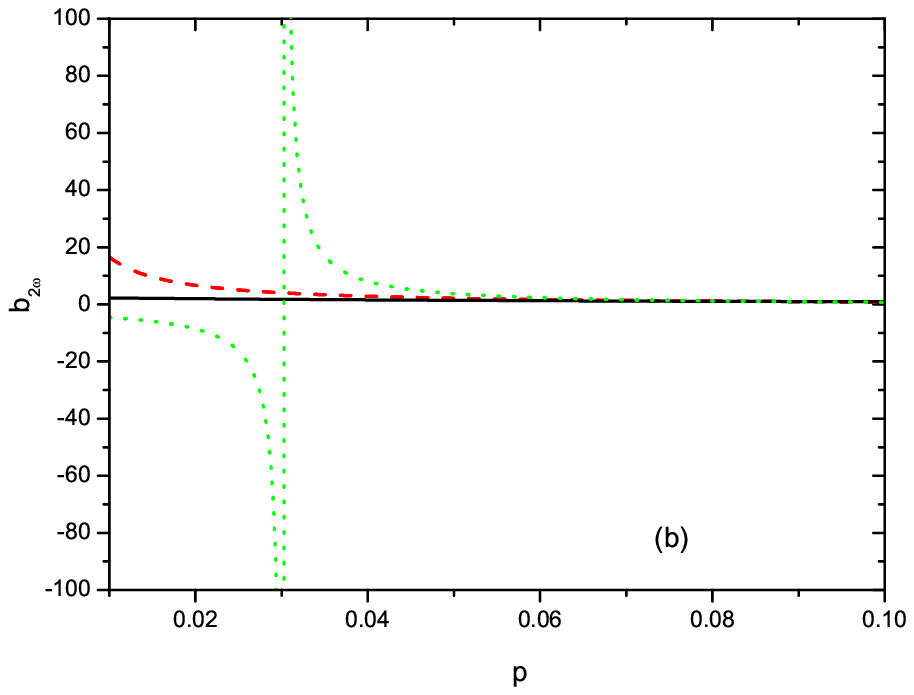}
\includegraphics[width=150pt]{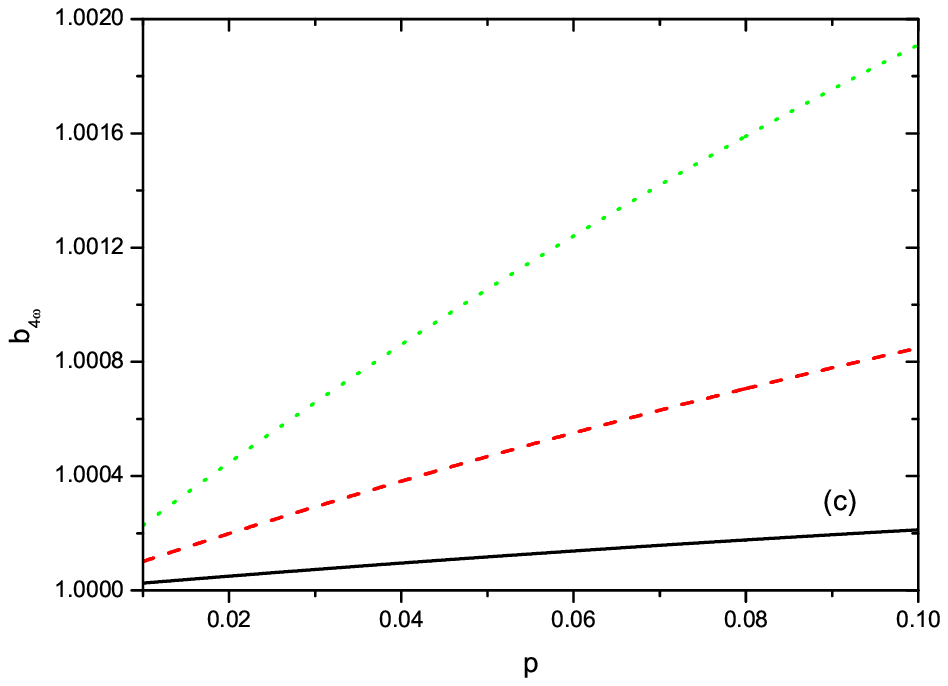}
\includegraphics[width=150pt]{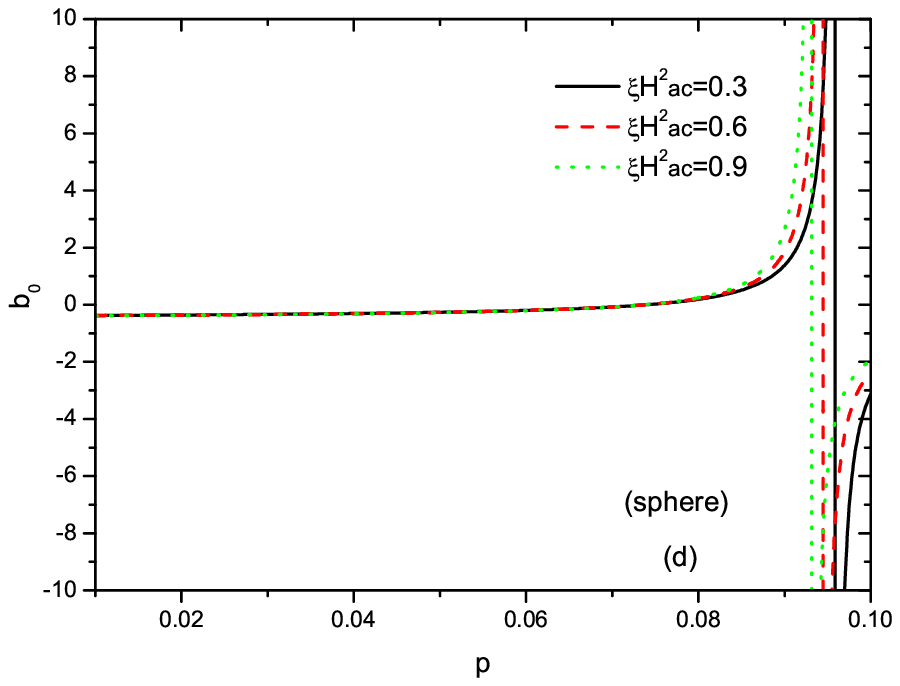}
\includegraphics[width=150pt]{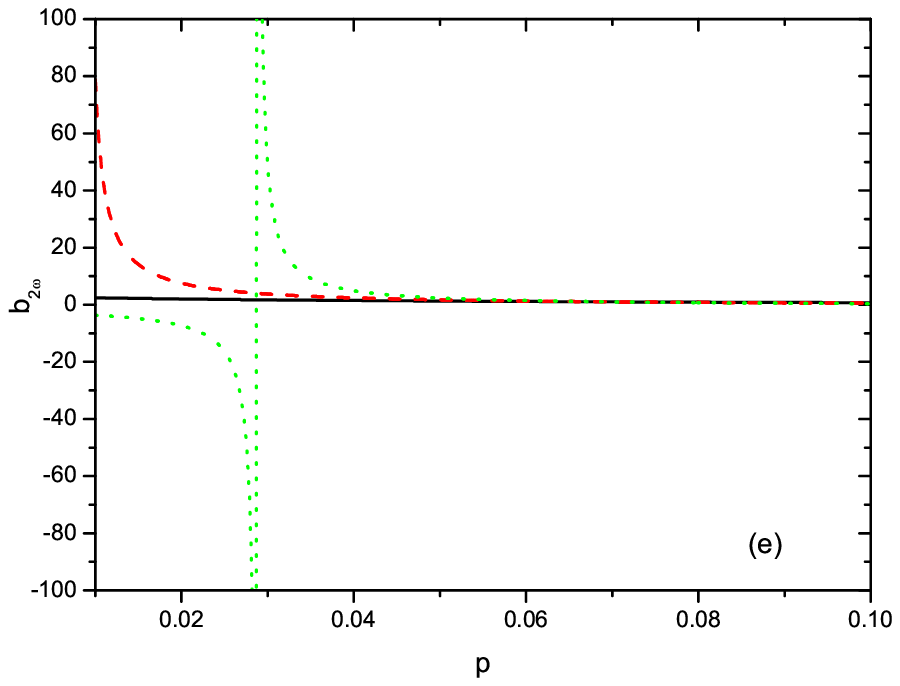}
\includegraphics[width=150pt]{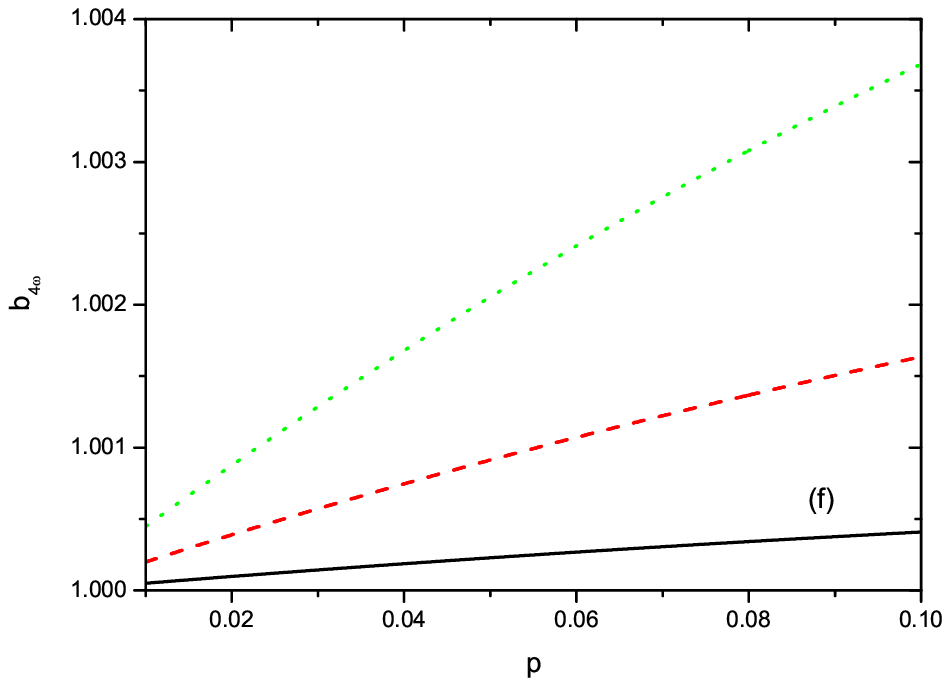}
\includegraphics[width=150pt]{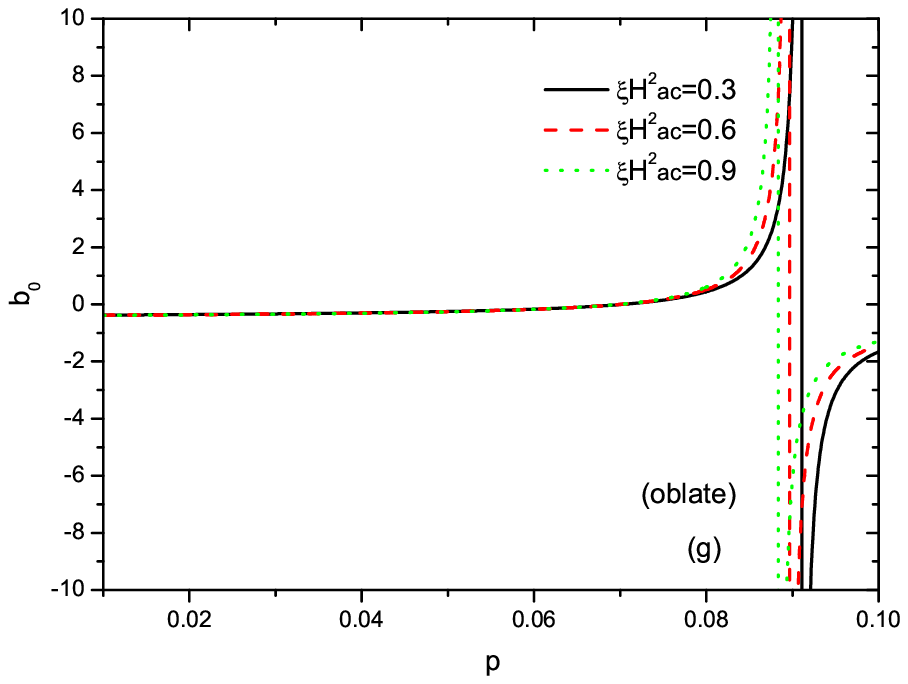}
\includegraphics[width=150pt]{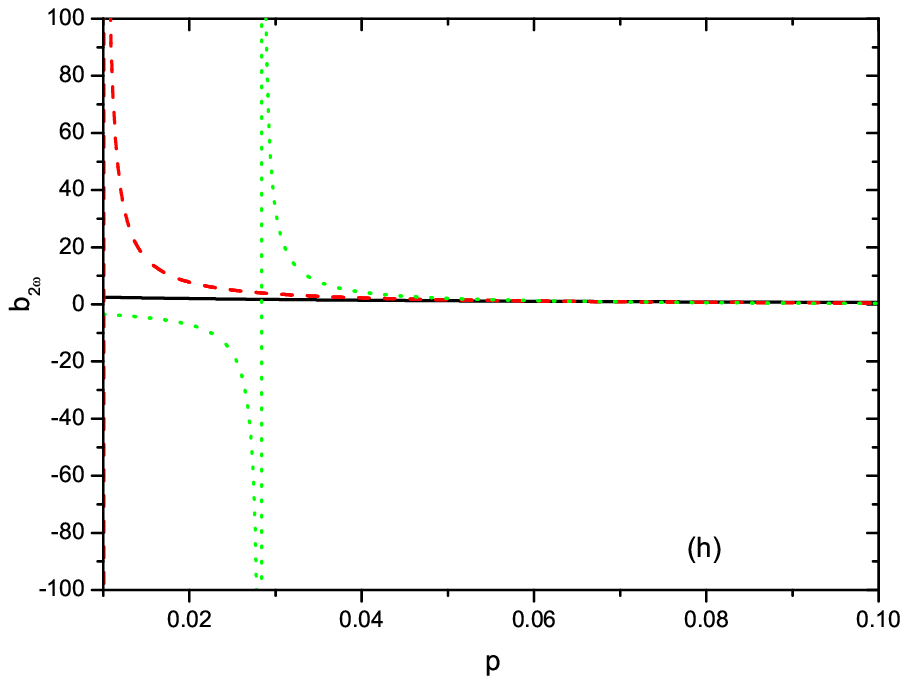}
\includegraphics[width=150pt]{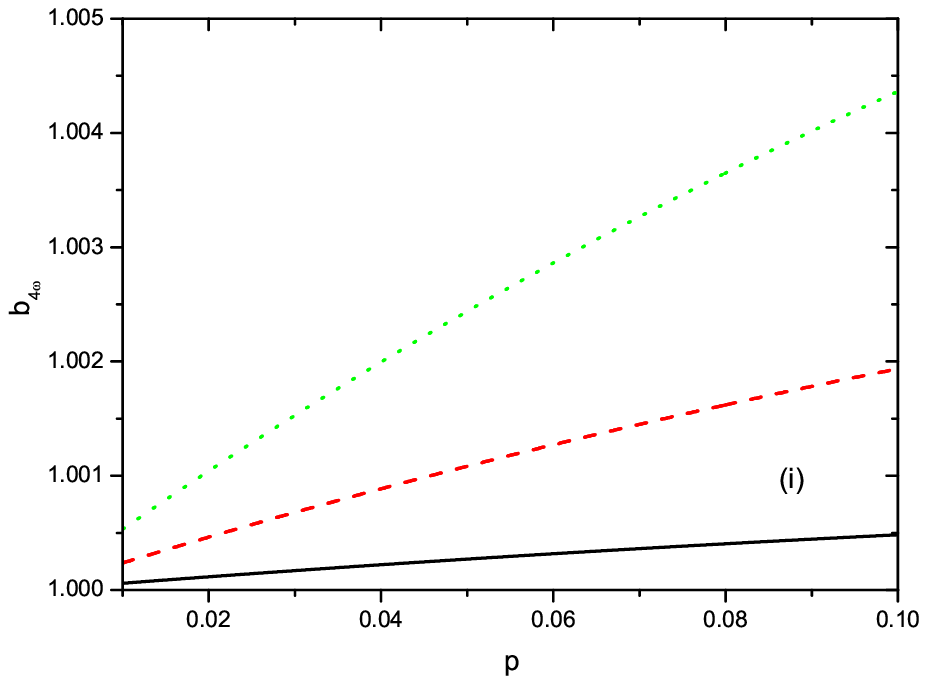}

\caption{Jian, Gao, and Huang}\label{}
\end{figure}
\end{document}